\documentclass{article}
\usepackage[a4paper, total={6in, 8in}]{geometry}
\usepackage{graphicx} 
\usepackage{slashed}
\usepackage{comment}
\usepackage{mdframed}
\usepackage[compat=1.0.0]{tikz-feynman}
\usepackage{appendix}
\usepackage{setspace}
\usepackage{hyperref}
\usepackage[scr=boondox]{mathalfa}
\usepackage{bbold}
\usepackage[backend=biber,style=phys,biblabel=brackets,sorting=none]{biblatex}
\newcommand{\be}{\begin{equation}}
\newcommand{\ee}{\end{equation}}
\newcommand{\bea}{\begin{eqnarray}}
\newcommand{\eea}{\end{eqnarray}}
\newcommand{\beas}{\begin{eqnarray*}}
\newcommand{\eeas}{\end{eqnarray*}}

\newcommand{\identity}{\mathbb{1}}
\newcommand{\RP}{{R}}
\newcommand{\PJ}{{P}}
\newcommand{\C}{{C}}

\begin{document}

\begin{titlepage}

\enlargethispage{0.5cm}
\vspace*{-1.5cm}

\begin{center}
{\Large Klein Bottle Cosmology }

\vspace{4mm}

\renewcommand\thefootnote{\mbox{$\fnsymbol{footnote}$}}

Brian Greene${}^{1}$\footnote{brian.greene@columbia.edu},
Daniel Kabat${}^{2,3}$\footnote{daniel.kabat@lehman.cuny.edu},
Janna Levin${}^{4}$\footnote{janna@astro.columbia.edu},
Massimo Porrati${}^{5}$\footnote{massimo.porrati@nyu.edu}

\vspace{1mm}

${}^1${\small \sl Departments of Physics and Mathematics, Columbia University} \\
{\small \sl 538 West 120th Street, New York, NY 10027, USA}


${}^2${\small \sl Department of Physics and Astronomy} \\
{\small \sl Lehman College, City University of New York} \\
{\small \sl 250 Bedford Park Blvd.\ W, Bronx, NY 10468, USA}


${}^3${\small \sl Graduate School and University Center, City University of New York} \\
{\small \sl  365 Fifth Avenue, New York, NY 10016, USA}


${}^4${\small \sl Department of Physics and Astronomy} \\
{\small \sl Barnard College of Columbia University} \\
{\small \sl New York, NY 10027, USA}


${}^5${\small \sl Center for Cosmology and Particle Physics} \\
{\small \sl Department of Physics, New York University} \\
{\small \sl 726 Broadway, New York, NY 10003, USA}

\end{center}


\noindent
We explore a higher-dimensional universe that is a product of Minkowski space and the nonorientable Klein bottle. The topology explicitly breaks important symmetries, such as translational invariance and $(5+1)$-dimensional $\C\PJ$ invariance. Somewhat surprisingly, the $(3+1)$-dimensional ${cp}$ of the Minkowski space can also be broken by the Klein bottle, both explicitly and in the presence of a brane. The topology enforces a background of fermion correlations that amounts to a condensate wall localized in the Klein bottle. The wall acts as an order parameter for the broken symmetries. If a brane passes through the wall, brane fermions that couple to the condensate are produced as quantified by the Bogoliubov coefficients for a time-dependent mass. The scenario meets the conditions, including ${cp}$ violation, to potentially generate the matter-antimatter asymmetry of the universe. 

\end{titlepage}

\setcounter{footnote}{0}
\renewcommand\thefootnote{\mbox{\arabic{footnote}}}

\tableofcontents 

\section{Extra Dimensions in Cosmology}

 The Klein bottle \cite{Klein1882} is an intriguing space to explore nonorientable extra dimensions, as detailed in a preceding paper \cite{Greene2025}. Notably, the Klein bottle induces condensate walls from bulk fields, a feature not seen in the torus, the orientable counterpart \cite{Greene2025,DeWitt_1979,DeWittMorette_1990}. 

In particular, a free, massless bulk fermion -- the minimal fermion with no new parameters introduced -- contributes a condensate wall with peaks around special axes: the flip axis at the origin of the twisted identification and the axis diametrically opposite the flip axis. 
Depending on the spin of the bulk field, the energy density of the wall may contribute to the cosmological constant, a possible signal of hidden dimensions \cite{Greene_2007,Greene2025}.

As well as generating condensates from bulk fields, the Klein bottle topology explicitly breaks symmetries on the covering space \cite{Greene2025}. Although the local geometry is flat and homogeneous, the topology explicitly breaks translational invariance, leading to a condensate wall that varies across the Klein bottle. The discrete symmetries $\C,\PJ,$ and $\C\PJ$ of the Dirac equation can also be broken.
A fermion transported once around the M\"obius direction of the Klein bottle flips chirality. There are various boundary conditions that can be imposed on a fermion under this reflection. Generally, they are either
periodic or antiperiodic  on the covering torus \cite{Witten2016}. Each of these boundary conditions explicitly breaks a different set of symmetries. The symmetry breaking is again manifest in the condensate.
Some boundary conditions will explicitly break charge symmetry and/or parity (and/or chirality). We review the explicitly broken symmetries found in \cite{Greene2025} and discuss the additional possibility of spontaneous symmetry breaking.

Perhaps most intriguing, $(5+1)$-dimensional $\C\PJ$ is explicitly broken, with various physical implications, including a possible source for the matter asymmetry in the universe. 
Curiously, $(3+1)$-dimensional ${cp}$ is also broken, which is significant if the condensate couples to a brane. For antiperiodic boundary conditions, $(3+1)$-dimensional ${cp}$ is explicitly broken while for periodic boundary conditions ${cp}$ is broken by the location of the brane. 

In this sequel to \cite{Greene2025,Greene_2011,Greene_2022,Greene_2023},
we explore the implications for the early universe of the broken symmetries in the Klein bottle. In particular, we imagine a $(3+1)D$ brane boosted along the Klein bottle and calculate particle production of $(3+1)D$ fermions as the brane passes through the condensate. The process we consider can lead to baryogenesis via leptogenesis. As a byproduct, we find a general expression for the Bogoliubov coefficients that quantify the particle production for a brane fermion with a time-dependent Dirac mass.

Further cosmological implications of the Klein bottle, which we leave for study elsewhere, include dark matter candidates that receive their mass from the condensate, dark energy contributions from the same, and a condensate-induced potential that can stabilize the volume of the 
Klein bottle as well as bring the brane to rest. 

\section{The Klein Bottle}

\subsection{Covering Space and $\Gamma$ Matrices}

We work in $(5+1)$ dimensions composed of a $(3+1)$ Minkowski spacetime times a $2D$ Klein bottle, ${\mathcal M}\times {\mathcal K}$.
We adopt a mostly minus metric $(+,-,-,-,-,-)$ for ease of comparison with canonical particle physics convention. The Dirac equation in $(5+1)$ dimensions is
$ (i\slashed D - m \identity_8)\psi=0$
with positive and negative frequency modes $\psi$ on the covering space.

We build $8\times 8$ $\Gamma^M$ matrices and a chirality operator $\bar \Gamma=\Gamma^0\Gamma^1\Gamma^2\Gamma^3\Gamma^4\Gamma^5$, using the notation $M=0,5$ and $J=1,5$. The $\Gamma$s obey the Clifford algebra
\begin{equation}
    \left \{ \Gamma^M,\Gamma^N\right \}=2\eta^{MN} \identity_8\quad \quad \quad \eta^{MN}={\rm diag}(1,-1,-1,-1,-1,-1) \  .
    \label{eq:cliff}
\end{equation}
It follows that 
$\left (\Gamma^{0}\right)^2=\identity_8$ and $\Gamma^0=\Gamma^{0\dagger}$ while
$\left (\Gamma^{J}\right)^2=-\identity_8$ and
$\Gamma^J=-\Gamma^{J\dagger}$. Also $\bar \Gamma=\bar \Gamma^\dagger$ and $\bar \Gamma^2=\identity_8$. We'll make use of $\{\Gamma^N,\bar \Gamma\}=0$.
The explicit representation is
\begin{eqnarray}
    \Gamma^\mu &=&\gamma^\mu \otimes \sigma^3 =\pmatrix{\gamma^\mu & 0 \cr 0 & -\gamma^\mu} \nonumber \\
    \Gamma^4 &=& \identity_4 \otimes i\sigma^1 =i\pmatrix{0 & \identity_4 \cr \identity_4 & 0} \nonumber \\
    \Gamma^5 &=&\identity_4 \otimes i\sigma^2  
    =\pmatrix{0 & \identity_4 \cr -\identity_4 & 0} 
    \nonumber \\
    \bar \Gamma &=&-\bar\gamma \otimes \sigma^3  
    =\pmatrix{
    -\bar\gamma & 0 \cr 0 & \bar\gamma}\ \ .
\end{eqnarray}
These are built with the $(3+1)$ $\gamma$ matrices which obey the Clifford algebra
$\left \{ \gamma^\mu,\gamma^\nu\right \}=2\eta^{\mu,\nu} \identity_4$ with
$\left (\gamma^{0}\right)^2=\identity_4$ and 
$\left (\gamma^{j}\right)^2=-\identity_4$.
With this convention $\gamma^0=\gamma^{0\dagger}$ and $\gamma^j=-\gamma^{j\dagger}$. 
In the chiral basis 
\begin{equation}
\gamma^\mu=\pmatrix{0& \sigma^\mu\cr \bar \sigma^\mu & 0}\quad  {\rm with}  \quad 
\sigma^\mu=\pmatrix{1_2 ,& \sigma^i} \quad {\rm and} \quad   \bar\sigma^\mu=\pmatrix{1_2 ,& -\sigma^i}
\ .
\end{equation}
Just to have on hand for convenience, the Hermitian Pauli matrices are
\begin{eqnarray}
    \sigma^1=\pmatrix{0 & 1 \cr 1 & 0} \quad \quad  \sigma^2=\pmatrix{0 & -i \cr i & 0}\quad \quad  \sigma^3=\pmatrix{1 & 0 \cr 0 & -1}
\end{eqnarray}
for which $\sigma^{i\dagger}=\sigma^i$, $(\sigma^i)^2=1_2$, $\{\sigma^i,\sigma^j\}=2\delta^{ij}\identity_2$ and $[\sigma^i,\sigma^j]=2i\epsilon_{ijk}\sigma^k$.
The chiral $\gamma$ will be called $\bar \gamma$ instead of the conventional $\gamma^5$ to avoid confusion with the fifth dimension. We have $\bar\gamma=i\gamma^0\gamma^1\gamma^2\gamma^3$ which is $ \bar\gamma={\rm diag}(-1,-1,1,1).$ Also $\bar\gamma=\bar\gamma^{\dagger}$ and $(\bar \gamma)^2=\identity_4$ and $\bar\gamma$ anticommutes with the others $\{\gamma^\mu,\bar\gamma\}=0$.

\subsection{Klein Bottle Boundary Conditions}

We build a Klein bottle by identifying the sides of a rectangle with a twist on traversing the $x_5$ direction
\begin{eqnarray}
    (x^\mu,x^4,x^5)
   & \approx&
    (x^\mu,x^4+2\pi r_4,x^5)\nonumber \\
    (x^\mu,x^4,x^5)
   & \approx&
    (x^\mu,-x^4,x^5+2\pi r_5)\ .
    \label{eq:BCKlein}
\end{eqnarray}
We introduce the notation for convenience,
\begin{equation}
    \tilde x=(x^\mu,-x^4,x^5+2\pi r_5)
    \label{eq:tildex}\ \ .
\end{equation}
The Klein bottle is nonorientable. To accommodate spinors in the Klein bottle, begin with the periodic modes on the covering torus \cite{AppelquistChodosFreund1987} of size $(2\pi r_4,2\pi (2r_5))$:
\begin{eqnarray}
\psi(x) &=& \psi(x^\mu,x^4+2\pi r_4,x^5)\label{eq:6b2p}\\
\psi(x) &=& \psi(x^\mu,x^4,x^5+2(2\pi r_5))
\ ,
\label{eq:6b3p}
\end{eqnarray}
of the form $\psi\sim e^{\mp ik\cdot x}$ (on-shell) with
\begin{equation}
k_4=\frac{n_4}{r_4}
\quad, \quad
k_5=\frac{n_5}{2r_5}\quad\quad\quad n_4,n_5\in  {\mathbb Z}
\ .
\label{eq:modes1}
\end{equation}
To determine which of these spinors descend to the Klein bottle, we need to specify the boundary conditions imposed in the twisted direction, of which there are a number of possibilities. Consider, for example, the reflection of a spinor upon being transported around $x_5$. We denote the reflection operation by $\RP_4$ and denote modes on the Klein bottle by $\Psi$ to distinguish from modes $\psi$ on the covering torus. The reflection matrix is 
\begin{equation}
    \RP_4=\Gamma^4\bar \Gamma
    \ ,
     \label{eq:R4}
\end{equation}
for which $\RP_4=\RP_4^\dagger$.
There is a notable phase ambiguity. The fermion can be periodic under reflection, which we will call $\RP_4^+$ boundary conditions,
\begin{eqnarray}
  \RP_4^+:\quad \Psi(x)= \RP_4\Psi(\tilde x)
    \ ,
    \label{eq:BCreq1}
\end{eqnarray}
using the notation of eqn.\ \ref{eq:tildex}. Periodicity is ensured since $\RP_4^2=\identity_8$, which on traversing $x_5$ twice gives $\Psi(x)=\Psi(x +4\pi r_5\delta^M_{5})$.  
The fermions can also be antiperiodic, which we call $\RP_4^-$ boundary conditions, 
\begin{eqnarray}
   \RP_4^-:\quad \Psi(x)= i\RP_4\Psi(\tilde x)
    \ ,
    \label{eq:BCreq1-}
\end{eqnarray}
Antiperiodicity is assured since $(i\RP_4)^2=-\identity_8$, which on traversing $x_5$ twice gives $\Psi(x)=-\Psi(x +4\pi r_5\delta^M_{5})$. 
For a more thorough discussion of general phases see \cite{Greene2025}.
We will be slightly cavalier with the notation and use $\RP_4$ to stand for both the operation and the matrix representation.

An additional possibility is that orientation reversal is accompanied by charge conjugation $\Psi^\C=\C\Psi^*$,
\begin{equation}
   \C\RP_4^+:\quad \Psi(x)=\C \RP_4\Psi^*(\tilde x) \ \ .
\end{equation}
(Note, we take $\C$ first then $\RP_4$, so $\C \RP_4=\RP_4\circ \C$.)
The charge conjugation matrix and $\C \RP_4$ are
\begin{eqnarray}
    \C&=&-i\Gamma^2\Gamma^4\bar \Gamma= -i\Gamma^2 \RP_4\label{eq:C} 
    \\
    \C\RP_4&=&-i\Gamma^2 \ \ .
\end{eqnarray}
There is again a phase ambiguity, and indeed we use a different phase than in our previous paper \cite{Greene2025}, which can be consulted for a more general discussion of phases. 
With these definitions, $\C^2=\identity$ and $\C=\C^\dagger$.
We take note that for any choice of phase 
\begin{eqnarray}
    (\Psi^\C)^\C=-\Psi \ \ ,
\end{eqnarray}
and the $\C\RP_4^+$ boundary conditions are always periodic on the covering torus.

The boundary conditions are satisfied by mode solutions of the form, due to the (anti)periodicity, 
\begin{eqnarray}
\RP_4^+:\quad \Psi(x)&=&\frac{1}{\sqrt{2}}\left (\psi(x)+\RP_4\psi(\tilde x)\right )\nonumber \\
\RP_4^-:\quad \Psi(x)&=&\frac{1}{\sqrt{2}}\left (\psi(x)+i\RP_4\psi(\tilde x)\right )\nonumber \\
\C\RP_4^+:\Psi(x)&=&\frac{1}{\sqrt{2}}\left (\psi(x)+\C\RP_4 \psi^*(\tilde x)\right )\ \ .\label{eq:BCs}
\end{eqnarray}
The full Kaluza-Klein (KK) spectrum of solutions was presented in \cite{Greene2025}. 

A note on chiral fermions. Since $\RP_4$ does not commute with $\bar\Gamma$, we are unable to construct chiral fermions. Indeed, the situation is a bit worse than the usual obstruction present in compactifications of higher-dimensional theories. For instance, the left or right-handed projection of a mode on the Klein bottle is not itself a solution for the reasons described above. Consequently,
bulk fermions cannot couple to a chiral gauge theory and must be singlets under the standard model gauge interactions. 
This would seem to preclude the 3 large dimensions from possessing nonorientable topology if the standard model gauge interactions are exact. To have chiral fermions, we therefore assume they are confined to a (3+1)D-brane that lives in the ${\cal M}\times {\cal K}$ spacetime.

\subsection{Explicitly Broken Discrete Symmetries}
\label{sec:brokensymm}

\begin{table}
    \centering
    \begin{tabular}{|c|c|c|c|c|c|c|}
    \hline
         &  &  &  &  &  &\\ 
          & \quad $\PJ_{123}$ \quad &\quad $\RP_4$ \quad & \quad $\RP_5$ \quad & \quad $\PJ$ \quad\quad &\quad $\C$ \quad \quad & \quad $e
         ^{i\alpha \bar \Gamma}$ \quad \\
         &  &  &  &  &  &\\
         \hline
         &  &  &  &  &  &\\
         $\RP_4^+:\ \Psi(x)=\RP_4\Psi(\tilde x)$& $-$ & $+$ &  $-$ & $+$ &  $-$  & $-$  \\
         &  &  &  &  &  &\\
         \hline
         &  &  &  &  &  &\\
         $\RP_4^-:\ \Psi(x)=i\RP_4\Psi(\tilde x)$& $-$ & $+$ & $+$ & $-$ & $+$ &  $-$ \\
         &  &  &  &  &  &\\
         \hline
         &  &  &  &  &  &\\
         $\C\RP_4^+:\ \Psi(x)=\C\RP_4\Psi^*(\tilde x)$& $-$ & $-$ & $+$ & $-$ & $-$ & $+$\\
         &  &  &  &  &  &\\
    \hline
    \end{tabular}
    \caption{The boundary conditions explicitly break some symmetries on the covering space. A $+$ indicates the symmetry is preserved, a $-$ that the symmetry is broken. We note that $\PJ_{123}=i\RP_1\RP_2\RP_3$ and so $\PJ=\Gamma^0=R_1R_2R_3R_4R_5=-iP_{123}R_4R_5$. In the final row, this phase difference accounts for the breaking of $\PJ$ but not $P_{123}R_4R_5$. In general, the final row is very dependent on phase choices in the definition of $\C$. The results for arbitrary phase are summarized in \cite{Greene2025}.}
    \label{tab:broke}
\end{table}

The fermion boundary conditions break some of the discrete symmetries we had on the covering space as summarized in table \ref{tab:broke} and detailed in \cite{Greene2025}. The massive Dirac equation is symmetric under $\C$ and $\PJ$ as well as under reflection in any given direction, $R_J=\Gamma^J\bar\Gamma$. We can ask if these transformations map solutions out of the solution space. 

For instance, with $R_4^+$ boundary condition,
\begin{eqnarray}
     \RP_4^+:\quad \Psi(x)=\RP_4\Psi(\tilde x)\ ,
\end{eqnarray}
 a Klein-bottle mode in the solution space is mapped under parity $\PJ$ to a new mode
\begin{eqnarray}
    \Psi^\prime(x)={\PJ} \Psi(t,-x^J)&=&\Gamma^0\Psi(t,-x^J)\nonumber\\
    &=&\Gamma^0\RP_4\Psi(t,-x^j,x^4,-x^5+2\pi r_5)\nonumber\\
    &=&\RP_4\Gamma^0\Psi(t,-x^j,x^4,-x^5+2\pi r_5)\nonumber\\
    &=&\RP_4\Psi^\prime(\tilde x)\ \ ,
\end{eqnarray}
from which we conclude that the parity transformed field obeys the boundary condition and therefore is also a solution.

However, for the $\C $ transformed field,
\begin{eqnarray}
    \Psi^\prime(x)=\C\Psi^*(x) &=&\C \RP_4^*\Psi^*(\tilde x)\nonumber\\
   &=&-\RP_4\C\Psi^*(\tilde x)\nonumber\\
    &=&-\RP_4\Psi^\prime(\tilde x)
    \nonumber\\
     &\ne &\RP_4\Psi^\prime(\tilde x)
    \ \ ,
\end{eqnarray}
from which we conclude that the $\C $ transformed field does not obey the boundary condition and therefore is not a solution. The $\RP_4^+$ boundary condition preserves $\PJ$ but explicitly breaks $\C$ as well as $\C\PJ$.

Following the same logic for $\RP_4^-$, we find $\PJ$ is explicitly broken while $\C$ is preserved. Again, the important combination $\C\PJ$ is broken \cite{Greene2025}.

For massless fermions, the Dirac equation is also chirally symmetric, under transformations of the field by $\bar\Gamma$,
$\Psi^\prime=e^{i\alpha\bar\Gamma}\Psi$.
We can ask if chiral symmetry for massless fermions is preserved. (More accurately, chiral rotations
when broken are broken to the ${Z}_2$ subgroup $\alpha \in \lbrace 0,\pi \rbrace$ for $\RP_4^\pm$.)
The results for the various boundary conditions and symmetries are tabulated in table \ref{tab:broke}. 

In \S \ref{sec:baryo}, we will be interested in coupling to a $(3+1)$D brane, so we also tabulate the outcomes for a $(3+1)$-dimensional parity, $\PJ_{123}=iR_1R_2R_3$. The phase is chosen so that $\PJ_{123}^2=\identity_8$.

\section{The Condensate Wall}

In the torus, the two-point correlator
encodes the free propagation of a fermion through space.
However, for the Klein bottle, the vacuum supports a self-consistent background of fermionic condensates in the absence of an external source. The correlator thereby acquires an additional, position-dependent expectation value even at coincident points. This nonzero expectation value is the fermion condensate, which we calculate here. 

\begin{figure}
    \centering{
    \includegraphics[width=0.35\linewidth]{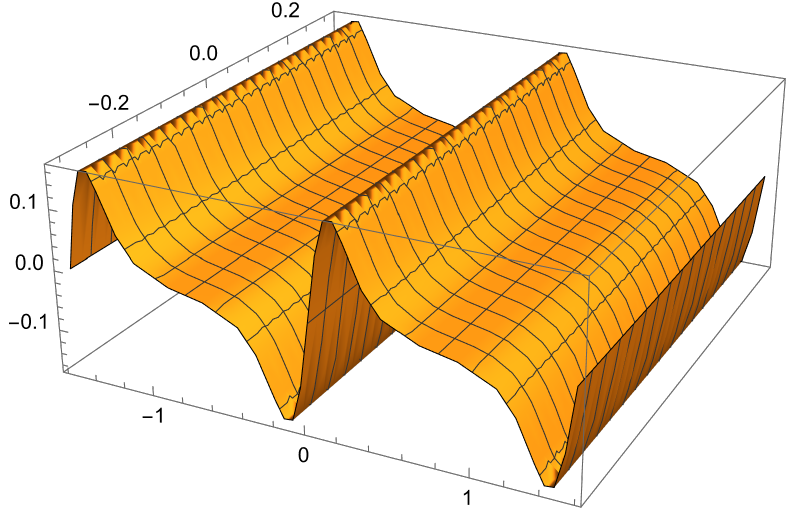}}
    \caption{The condensate wall as a function of $x_4$ and $x_5$.}
    \label{fig:wall6D}
\end{figure}

To keep the scope manageable, we will work with $\RP_4^+$ boundary conditions. (The correlators were calculated for all three boundary conditions using different methods in \cite{Greene2025}.)
To isolate the condensate, we express the correlator in terms of the free propagator and the condensate wall:
\begin{eqnarray}
    \langle \Psi(x) \bar\Psi(x^\prime)\rangle=S_{T^2}(x,x^\prime)+i\bar\Gamma W(x,x^\prime)\ \ ,
\end{eqnarray}
where our notation exploits the fact that the free propagator is the same as the free propagator on the torus and the wall $W$ is the unique contribution to the condensate. This decomposition and the $\Gamma$ dependence are a bit of foreshadowing, as is the depiction in figure \ref{fig:wall6D}.

In addition to $\tilde x$ defined in eqn.\ (\ref{eq:tildex}), we introduce the notation
\begin{eqnarray}
    \tilde{\tilde x} &=& (x^\mu,-x^4,x^5-2\pi r_5)\nonumber \\
    \tilde k &=& (k^\mu,-k^4,k^5)\ \ ,
\end{eqnarray}
for which we have the useful combinations
\begin{eqnarray}
    k\cdot (\tilde x-\tilde x^\prime) &=& \tilde k\cdot ( x-x^\prime)
\nonumber \\
k\cdot ( \tilde x-x^\prime)&=&   \tilde  k\cdot (x-\tilde{\tilde x}^\prime)
\ \ .
    \label{eq:combs}
\end{eqnarray} 
We will consider $\RP_4^+$ boundary conditions for which the modes can be decomposed as
\begin{eqnarray}
\Psi(x)&=&\frac{1}{\sqrt{2}}\left (\psi(x)+\RP_4 \psi(\tilde x)\right )\nonumber \\
\bar \Psi(x^\prime) &=& \frac{1}{\sqrt{2}}\left (\bar \psi(x^\prime) +\bar \psi(\tilde x^\prime) \RP_4^\dagger\right )
\ \ .
\label{eq:PBC}
\end{eqnarray}
The condensate in terms of modes on the covering torus is then
\begin{eqnarray}
   2\left < \Psi(x) \bar \Psi(x^\prime) \right > =\left < \psi(x) \bar \psi(x^\prime) \right >&+&\RP_4\left < \psi(\tilde x) \bar \psi(\tilde x^\prime) \right >\RP_4^\dagger\nonumber \\&+&\left < \psi(x)\bar \psi(\tilde x^\prime) \right > \RP_4^\dagger +\RP_4\left < \psi(\tilde x)\bar \psi(x^\prime) \right > \ \ .\nonumber
    \end{eqnarray}
The first term is the propagator on the covering torus. 
The second term is
\begin{eqnarray}
    \RP_4\left < \psi(\tilde x) \bar \psi(\tilde x^\prime) \right >\RP_4^\dagger&=& \sum_{w_4}\sum_{w_5}
   \int  \frac{d^6k}{(2\pi)^6}
   \frac{i \RP_4(\slashed { k}+m) \RP_4^\dagger}{k^2-m^2+i\epsilon}
   e^{-ik\cdot (\tilde x-\tilde x^\prime)}e^{-ik_42\pi w_4r_4}e^{-ik_5 4\pi w_5r_5}
   \nonumber \\
    &=& \sum_{w_4}\sum_{w_5}
     \int  \frac{d^6k}{(2\pi)^6}
   \frac{i (\slashed {\tilde k}+m) }{k^2-m^2+i\epsilon}
   e^{-i\tilde k\cdot ( x- x^\prime)}e^{-ik_42\pi w_4r_4}e^{-ik_5 4\pi w_5r_5}\nonumber \\
    &=& \sum_{w_4}\sum_{w_5}
     \int  \frac{d^6k}{(2\pi)^6}
   \frac{i (\slashed { k}+m) }{k^2-m^2+i\epsilon}
   e^{-ik\cdot ( x- x^\prime)}e^{-ik_42\pi w_4r_4}e^{-ik_5 4\pi w_5r_5}\ \ ,\nonumber
\end{eqnarray}
having used $\RP_4\slashed k \RP_4^\dagger=\slashed{\tilde k}$ and $\RP_4\RP_4^\dagger=\identity_8$. In the last step we redefine $k_4\rightarrow -k_4$ and $w_4\rightarrow -w_4$. This conspires with the first term to give the free propagator for $\langle \Psi(x)\bar\Psi(x^\prime)\rangle $, which matches that on the covering torus, $S_{T^2}(x,x^\prime)$. We hereafter drop this toroidal contribution, which will vanish on taking the trace in the bilinear built from the condensate, and focus on the new contribution from the Klein bottle.
For the two last terms in $2\left < \Psi \bar \Psi \right >$, we have $\left < \psi(x)\bar \psi(\tilde x^\prime) \right > \RP_4^\dagger +\RP_4\left < \psi(\tilde x)\bar \psi(x^\prime) \right >$
\begin{eqnarray}
       & =&  \sum_{w_4}\sum_{w_5}\left [
   \int  \frac{d^6k}{(2\pi)^6}
   \frac{i  }{k^2-m^2+i\epsilon}
\left ((\slashed { k}+m) e^{-i k\cdot ( x- \tilde x^\prime)}\RP_4^\dagger
   +\RP_4(\slashed k+m)
   e^{-i k\cdot (\tilde  x-  x^\prime)}
   \right )
   \right ]e^{-ik_42\pi w_4r_4}e^{-ik_5 4\pi w_5r_5}\nonumber \\
   & =&  \sum_{w_4}\sum_{w_5}\left [
   \int  \frac{d^6k}{(2\pi)^6}
   \frac{i  }{k^2-m^2+i\epsilon}
\left ((\slashed { k}+m) e^{-i k\cdot (  x- \tilde x^\prime)}
   +(\slashed {\tilde k}+m)
   e^{-i \tilde k\cdot ( x-  \tilde{\tilde x}^\prime)}
   \right )\RP_4
   \right ]e^{-ik_42\pi w_4r_4}e^{-ik_5 4\pi w_5r_5}
   \nonumber \\
   & =&  \sum_{w_4}\sum_{w_5}\left [
 \int  \frac{d^6k}{(2\pi)^6}
   \frac{i  }{k^2-m^2+i\epsilon}
\left ((\slashed { k}+m) e^{-i  k\cdot (  x- \tilde x^\prime)}
   +(\slashed k+m)
   e^{-i k\cdot (  x-  \tilde{\tilde x}^\prime)}
   \right )  \RP_4
   \right ]e^{-ik_42\pi w_4r_4}e^{-ik_5 4\pi w_5r_5}
   \nonumber \\
   & =&  (i\slashed D +m)\RP_4\sum_{w_4}\sum_{w_5}\left [
 \int  \frac{d^6k}{(2\pi)^6}
   \frac{i  }{k^2-m^2+i\epsilon}
\left ( e^{-i  k\cdot (  x- \tilde x^\prime)}
   +
   e^{-i k\cdot (  x-  \tilde{\tilde x}^\prime)}
   \right )  
   \right ]e^{-ik_42\pi w_4r_4}e^{-ik_5 4\pi w_5r_5}
   \nonumber
   \end{eqnarray}
   where $\RP_4^\dagger=\RP_4$ and we have used eqn.\ (\ref{eq:combs}).
   Since the sums are over all $w_5$, we can shift to $w_5\rightarrow -w_5$ with no affect on the second term in parentheses. Then,
to consolidate the expression, we write
   \begin{eqnarray}
     2\left < \Psi (x)\bar\Psi(x^\prime) \right > &=& (i{\slashed D }+m)\RP_4 \sum_{w_4}\sum_{w_5}
   \left (D_F( x- \tilde X^\prime)+D_F( x- \tilde{\tilde X}^\prime) \right )
   \ \ ,
   \label{eq:Pcon2tilde}
\end{eqnarray}
where 
$D_F$ is the $(5+1)D$ scalar propagator and we define
\begin{eqnarray}
        \tilde X^\prime &= & (x^{\prime \mu},-x^{\prime 4}+2\pi w_4r_4,x^{\prime 5}+2\pi (2w_5+1) r_5)\nonumber \\
         \tilde{\tilde X}^\prime &=&
         (x^{\prime \mu},-x^{\prime 4}+2\pi w_4r_4,x^{\prime 5}-2\pi (2w_5+1) r_5)
         \ \ .
         \label{eq:xxs}
\end{eqnarray}
In the coincident limit $x\rightarrow x^\prime$, only the $\partial_{4,5}$ terms
survive, giving
\begin{eqnarray}
2\left < \Psi (x)\bar\Psi(x^\prime) \right >&=& 
  \sum_{w_4}\sum_{w_5}(i\Gamma^4{\partial_4}+i\Gamma^5\partial_5+m)\RP_4
\left (D_F(x-\tilde X^\prime)+D_F(x-\tilde{\tilde X}^\prime)\right)
 \nonumber \\
&=&
 i \sum_{w_4}\sum_{w_5}
 \left(\Gamma^4\frac{(x_4-\tilde X_4^\prime)}{r}+\Gamma^5\frac{(x_5-\tilde X_5^\prime)}{r}\right)\RP_4\partial_r 
   D_F(x-\tilde X^\prime)\nonumber \\
   &\ &\quad\quad
 i \sum_{w_4}\sum_{w_5}\left(\Gamma^4\frac{(x_4-\tilde{\tilde X}_4^\prime)}{r}+\Gamma^5\frac{(x_5-\tilde{ \tilde X}_5^\prime)}{r}\right)\RP_4\partial_r 
   D_F(x-\tilde{ \tilde X}^\prime)
   \nonumber \\
   &\ &\quad\quad\quad
   + \sum_{w_4}\sum_{w_5} m \RP_4
   \left 
(D_F(x-\tilde X^\prime)+D_F(x-\tilde{\tilde X}^\prime)\right)\ \ .
\end{eqnarray}
In the limit of $x^\prime\rightarrow x$, $(x_5-\tilde{\tilde X}_5^\prime)=2\pi (w_5+1)r_5=-(x_5-{\tilde X}_5^\prime)$ and the terms proportional to $\Gamma^5$ cancel. 
Going over to the massless case to make this simpler
\begin{eqnarray}
    D_F(x-\tilde X^\prime)&=&\int  \frac{d^6k}{(2\pi)^6}
   \frac{i  }{k^2+i\epsilon}
e^{-ik\cdot (  x- \tilde X^\prime)}
=\frac{1}{4\pi^3}\left (\frac{1}{ (x-\tilde X^\prime)\cdot (x-\tilde X^\prime)}\right )^2\ \ .
\label{eq:DFspace}
\end{eqnarray}
In the limit $x^\prime\rightarrow x$, $D_F(x-\tilde X^\prime)=D_F(x-\tilde{\tilde X}^\prime)=D_F(r)$:
\begin{eqnarray}
D_F(r) &=&
\frac{1}{4\pi^3}\frac{1}{r^4}\nonumber \\
    r^2 &=& (2x_4-2\pi w_4r_4)^2+(2\pi (w_5+1)r_5)^2 \ \ .
    \label{eq:rs}
\end{eqnarray}
We then have,
\begin{eqnarray}
\left < \Psi \bar \Psi (x)\right >
   &=&-\frac{i}{\pi^3}\Gamma^4\RP_4 \sum_{w_4}\sum_{w_5}
  \frac{(2 x_4-2\pi w_4r_4)}{ r^6}
  \ \ .
    \label{eq:Pcon1}
\end{eqnarray}
We consolidate these results as 
\begin{mdframed}
\begin{eqnarray}
 \RP_4^+:\quad (m=0)\quad \left < \Psi \bar \Psi (x)\right >
   =-iW(x_4)\Gamma^4\RP_4 =iW(x_4)\bar \Gamma
\label{eq:cond6D}
\end{eqnarray}
\end{mdframed}
with the wall coefficient defined as
\begin{mdframed}
\begin{eqnarray}
    W(x_4)  =\frac{1}{\pi^3}
    \sum_{w_4}\sum_{ w_5}
    \frac{
       (2x_4-2\pi w_4r_4)
 }{\left (
 (2x_4-2\pi w_4r_4)^2+(2\pi (2w_5+1)r_5)^2
\right )^3}\ \ .
\end{eqnarray}
\end{mdframed}

The condensate is forced to zero at $x_4=0$ because the sum is odd in $w_4$. The condensate is also forced to zero on the edges, $x_4=\pm \pi r_4$, as can be seen by shifting to $w^\prime_4=w_4\mp 1$, respectively, so that the sum will be odd and exactly zero on those axes.

\begin{figure}
    \centering{
\includegraphics[width=0.4\linewidth]{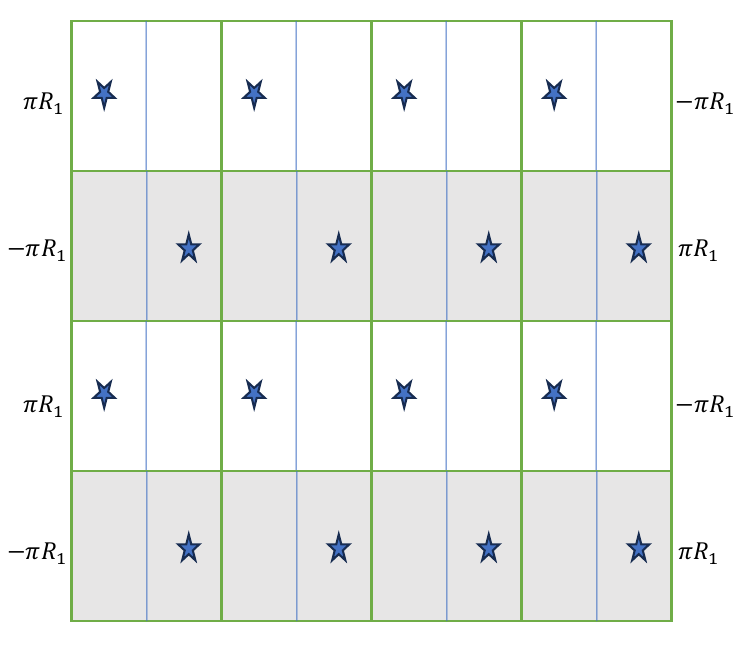}
\includegraphics[width=0.425\linewidth]{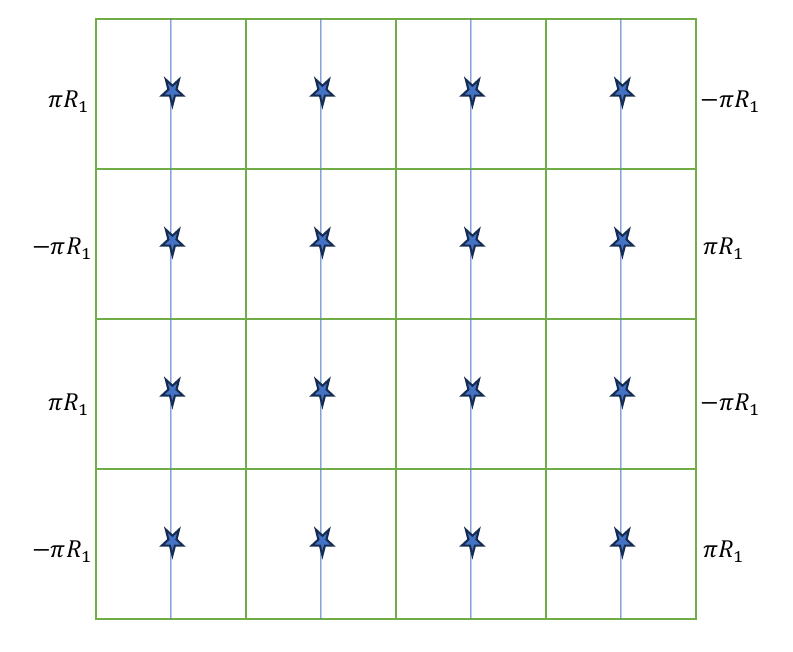}
}
\caption{Klein bottle tilings of the plane. The horizontal axis is $x_4$ and the vertical axis is $x_5$. The flip axis is indicated by the faint vertical blue line bisecting the tiles. Left: If a star is arbitrarily off the flip axis, there are a series of images in a $T^2$ of size $2\pi (r_4,2r_5)$ and there is a second series of images in all the shaded cells at $-x_4$ as well as all of its images in a $T^2$ of size $2\pi(r_4,2r_5)$. Right: For a star located on the flip axis, all images look like those in a $T^2$ of size $2\pi(r_4,r_5)$. The same would be true for the images of a star located at the identified points $x_4=\pm \pi r_4$. We find the condensate vanishes exactly on, and is centered around, these two special axes: the flip axis at $x_4=0$ and the identified axes $x_4=\pm \pi r_4$.}
        \label{fig:tiles}
\end{figure}
As shown in figure \ref{fig:wall6D}, there are two double-bumped walls around these axes, $x_4=0$ 
and $x_4=\pm \pi r_4$,
which are special locations where the infinite images are those of a simple torus and translational symmetry is restored.\footnote{In other words, translations in the
$x^4$ direction
\begin{equation}
    \Psi'(x) = \Psi(x^\mu,x^4 + a, x^5)
\end{equation}
are broken to the ${Z}_2$ subgroup $a \in \lbrace 0, \pi r_4 \rbrace$.}
Fig. 2 depicts the pattern of images for the Klein bottle and also shows that the pattern of images reduces to those for a torus on these special axes.
The modes $w_4=-1,0,1$ capture the dominant features of both hills. 
Including both walls to lowest order gives
\begin{eqnarray}
 W(x_4)&\approx&\frac{1}{\pi^3}\left [
    \frac{   (2x_4) }{\left (
 (2x_4)^2+(2\pi r_5)^2
\right )^3}+
    \frac{
       (2x_4\mp 2\pi r_4)
 }{\left (
 (2x_4\mp 2\pi r_4)^2+(2\pi r_5)^2
\right )^3}
\right ]\label{eq:wall6D}
\end{eqnarray}
(which is odd and vanishes on integration over the Klein bottle). 
The first term has a maximum at $x_{4max}=\frac{\pi}{\sqrt{5}}r_5$, so we'll take $r_4>\frac{\pi}{\sqrt{5}}r_5$ for the sake of argument. In figure \ref{fig:wall6D}, we show the wall for $r_4=4r_5$. As $r_4\gg r_5$, the distance between the peaks at the origin and those on the $x_4=\pm \pi r_4$ edges becomes longer and flatter.

\begin{table}
    \centering
    \begin{tabular}{|c|c|c|c|c|}
    \hline
         &  &  &  &\\ 
         B.C. & Condensate &  Value & Bilinear & $\langle$Bilinear$\rangle$\\
         &  &  &  &\\
         \hline
         &  &  &  &\\ 
         $\RP_4^+$& $\left <\Psi \bar\Psi\right >$ & $-iW(x_4)\Gamma^4\RP_4$ & $\bar\Psi i\bar \Gamma \Psi$ & $8W(x_4)$\\
         &  &  &  &\\
         \hline
         &  &  &  &\\ 
         $\C\RP_4^+$& $\left <\Psi \bar\Psi^\C\right >$ & $-iW(x_4)\Gamma^4\RP_4$ & $\bar\Psi^\C i\bar \Gamma \Psi$ & $8W(x_4)$\\
         &  &  &  &\\
         \hline
    \end{tabular}
    \caption{The first three columns are the boundary condition, the corresponding condensate, and its explicit value. The condensate has the same symmetries tabulated in table \ref{tab:broke}. }
    \label{tab:bilinearP4}
    \end{table}

The pseudoscalar bilinear with a nonzero vev for the condensate from the $\RP_4^+$ boundary conditions is
\begin{eqnarray}
    \left <\bar \Psi i\bar\Gamma\Psi \right >=-i{\rm Tr}(\bar\Gamma\left <\Psi \bar \Psi\right >)=W(x_4){\rm Tr}(\bar\Gamma\bar\Gamma)=8W(x_4)
\end{eqnarray}
as summarized in table \ref{tab:bilinearP4}. As expected, the condensate respects the same symmetries that the boundary conditions do in table \ref{tab:broke}. All other attempts to construct scalars or pseudoscalars would vanish on taking the trace. All toroidal contributions also vanish on taking the trace.

Similarly, from $\C\RP_4^+$ boundary conditions, a pseudoscalar bilinear with a nonvanishing vev,
$\left <\bar \Psi^\C i\bar\Gamma\Psi \right >= 8W(x_4) $, 
can be constructed from the condensate $\left < \Psi \bar \Psi^\C \right >$ with $\bar \Psi^\C=\Psi^T\C^\dagger\Gamma^0 $.
As above, all other attempts to construct scalars or pseudoscalars would vanish on taking the trace. 

The bilinear and condensate wall for $\RP_4^-$ boundary conditions are found in \cite{Greene2025}. It is interesting that the periodic boundary conditions lead to a wall that is antisymmetric in $x_4$. Conversely, the antiperiodic boundary conditions lead to a wall that is symmetric in $x_4$.

\section{Leptogenesis and Baryogenesis}
\label{sec:lepto}

The condensates can be put to good use to explain the excess of matter over antimatter. By introducing a 3D brane moving inside the Klein bottle and coupling a brane fermion to the bulk fermion, we can meet Sakharov's three conditions for successful baryogenesis \cite{Sakharov1967}: 1) Baryon number violation 2) ${cp}$ violation, and 3) out-of-equilibrium dynamics.

As the brane moves through the condensate walls, the mass term varies sharply, which leads to the production of brane fermions as we show explicitly in \S \ref{sec:bogo}. The production mechanism is a strictly out-of-equilibrium condition. Additionally, there can be lepton number violating mass terms so that leptogenesis can accompany the bursts of particle production. The standard model on the brane then provides the baryon number violation through $B-L$ conserving processes. The location of the brane breaks translation symmetry, which in turn leads to  spontaneous breaking of certain ${cp}$ symmetries, as we discuss below.

\subsection{Brane Fermion Masses}

We consider brane-bulk couplings in which the bulk condensate provides a position-dependent mass to brane fermions.
The standard definitions for charge conjugation ${c}$ and parity ${p}$ 
in 3+1 dimensions are
\begin{eqnarray}
    \nonumber
{c} \, : \, f(x) & \rightarrow & -i\gamma^2 f^*(x) \\
\label{eq:4Dparity}
{p} \, : \, f(x) & \rightarrow & \gamma^0 f(t,-x^1,-x^2,-x^3)
\end{eqnarray}

The properties of the bilinear $\bar \Psi i\bar\Gamma  \Psi $ are summarized in table \ref{tab:biCP}. The properties of various bilinears in $f$ are summarized in table \ref{tab:bilinearf}. Finally, the
properties of various mass terms are summarized in table \ref{tab:inter}.

\begin{table}[h]
    \centering
  \begin{tabular}{|c|c|c|c|c|c|}
        \hline 
        & & & & &\\
        B.C. & Bilinear& $\C\PJ$ &$\C\PJ_{123}$ &   $\C\PJ_{1234}$ & $\C\PJ_{1235}$ \\ 
        & & & & &\\
        \hline
        & & & & &\\
        $\RP_4^+$ & $ \bar \Psi i\bar\Gamma  \Psi  $ &  $\times$ & $ \bar \Psi i\bar\Gamma  \Psi |_{-\vec x} $  & $ -\bar \Psi i\bar\Gamma  \Psi |_{-\vec x,-x^4} $ & $\times$   \\
         & & & & &\\
        \hline
    \end{tabular}
    \caption{Transformation of the $\RP_4^+$ bilinear under various charge-parity transformations. These are ${\C\PJ=\C}{\Gamma^0}$, $\C\PJ_{123}=\C i\RP_1\RP_2\RP_3 $,  $\C\PJ_{1234}=\C\PJ_{123}\RP_4 $ and $\C\PJ_{1235}=\C\PJ_{123}\RP_5 $. The symbol $\times$ indicates an explicit breaking of the symmetry by the boundary conditions.}
    \label{tab:biCP}
    \hspace{2pt}

       \begin{tabular}{|c|c|c|c|}
        \hline 
        & & & \\
        $\bar f f$ & $\bar f i\bar\gamma f$ & $(\bar f^{c} f+\bar f f^{c})$ & $ (\bar f^{c} f-\bar f f^{c})$  \\ 
        & & & \\
        \hline
        & & & \\
        $+$ & $- $ &  $+$ & $- $  \\
         & & & \\
        \hline
    \end{tabular}
    \caption{$(3+1)$-dimensional ${cp}$ transformation of brane bilinears in $f$.}
    \label{tab:bilinearf}
    \hspace{2pt}
    \end{table}
    
Of these, the most interesting is the Majorana interaction 
\begin{eqnarray}
     {\cal L}_{\rm Maj}=-\frac{ig}{2}\left (\bar f^{c} f-\bar f f^{c}\right ) \left ( \bar \Psi i\bar\Gamma  \Psi  \right )\ \ ,
\end{eqnarray}
where $g$ is real. The term allows for lepton number violation, which is needed for baryogenesis. 
The interaction is ${cp}$ even under $f\rightarrow f^{{cp}}$ when combined 
with the $\C\PJ_{1234}=\C\PJ_{123}\RP_4 $ symmetry. 
 
This symmetry is spontaneously broken by the location of the brane, as is evident from the mass term generated by the vev of the bilinear,
\begin{eqnarray}
     {\cal L}_{\rm Maj}
     \rightarrow
    \frac{1}{2}\left (\bar f^{c} f-\bar f f^{c}\right ) im_f
    \ \ ,
\end{eqnarray}
which is odd under $f\rightarrow f^{{cp}}|_{-\vec x}$.
The real parameter
\begin{eqnarray}
 m_f=
 g
\langle\bar\Psi
\bar\Gamma \Psi\rangle=-ig{\rm Tr}\left (\bar \Gamma \langle\Psi\bar\Psi\rangle\right )
    =8gW(x_4)  
    \label{eq:mf}
\end{eqnarray}
leads to the imaginary mass $im_f$.
If the brane is at $x_4=0$ (or the identified edges $x_4=\pm \pi r_4$), then the mass vanishes and the symmetry is restored. If the brane is not located on these two special axes, then the mass term appears to a brane observer to be ${cp}$ violating.
This mass term now introduces two of the key elements needed for leptogenesis, lepton number violation and ${cp}$ violation.

    \begin{table}[h]
    \centering
  \begin{tabular}{|c|c|c|c|c|c|}
        \hline 
        & & & & &\\
        Interaction & $\Delta L$ & ${cp}+\C\PJ_{123}$ &   ${cp}+\C\PJ_{1234}$ & mass term & ${cp}$ \\ 
        & & & & &\\
        \hline
        & & & & &\\
         $-g\bar f  f \left ( \bar \Psi i\bar\Gamma  \Psi  \right )$ & no & $+$ & $-$ & $-\bar f f m_f$ &  $+$ \\
         & & & & &\\
        $-g\bar f i\bar \gamma f \left ( \bar \Psi i\bar\Gamma  \Psi  \right )$ & no & $-$ & $+$ &$-\bar f i \bar \gamma f m_f$ & $-$ \\
         & & & & &\\
        $-\frac{g}{2}\left (\bar f^{c} f+\bar f f^{c}\right ) \left ( \bar \Psi i\bar\Gamma  \Psi  \right )$ & yes & $+$ & $-$ & $-\frac{1}{2}\left (\bar f^{c} f+\bar f f^{c}\right ) m_f$ & $+$ \\
        & & & & &\\
        $-\frac{ig}{2}\left (\bar f^{c} f-\bar f f^{c}\right ) \left ( \bar \Psi i\bar\Gamma  \Psi  \right )$ & yes & $-$ & $+$ & $-\frac{1}{2}\left (\bar f^{c} f-\bar f f^{c}\right ) im_f$ & $-$ \\
    
        & & & & &\\
        \hline
    \end{tabular}
    \caption{Spontaneous $(3+1)$-dimensional ${cp}$ violation. Various Hermitian interactions between a brane fermion and a bulk fermion subject to $\RP_4^+$ boundary conditions on the Klein bottle are shown. Each interaction leads to a mass term for the brane fermion through the condensate. Both $g$ and $m_f=8W(x_4)$ are real. The columns tabulate lepton number violation; the interaction’s transformation under ${cp} + \C\PJ_{123}$; the interaction’s transformation under ${cp} + \C\PJ_{1234}$; the mass term from the vev of the bulk bilinear; and finally the subsequent ${cp}$ character of the mass term. A $+$ indicates the symmetry is preserved and a $-$ that the symmetry is broken. The second and fourth rows have a ${cp} + \C\PJ$ invariance that is spontaneously broken by the brane.}
    \label{tab:inter}
\end{table}

\subsection{Particle production from brane motion through the wall}
\label{sec:bogo}

If the brane is boosted in the $4$-direction with position $x_{b4}=v_4 t$, then as the brane moves through the condensate wall there will be bursts of particle production on the sharp slopes near $x_4\sim 0$ and $x_4\sim \pm \pi r_4$. 
Although above we consider a variety of masses, here we will just treat $f$ as a solution to the ordinary Dirac equation with a Dirac mass in order to streamline the calculation for the particle production. We'll simplify by imposing $r_4 \gg r_5$ and just consider scattering through the wall around the origin for which
\begin{eqnarray}
    m_f=\frac{8g}{\pi^3}\left (
  \frac{2{v_4 t}}{r^6}\right )
\end{eqnarray}
with $r^2=(2x_4)^2+(2\pi r_5)^2$ from eqn.\ (\ref{eq:rs}) with $w_4=w_5=0$.
We expect a maximum of $m_f$ at $x_4=\pi r_5/\sqrt{5}$, as shown in Fig. 3. 

To begin, we expand in
the Fourier modes $f_k$:
    \begin{eqnarray}
    f(x)&=&\int \frac{d^3k}{(2\pi)^3}e^{i{\mathbf k}\cdot {\mathbf x}}\sum_s \left (a_{{\mathbf k},s}f^u_{{\mathbf k},s}+b^\dagger_{-{\mathbf k},s}f^v_{-{\mathbf k},s}
    \right )
    \ \ ,
    \label{eq:fk}
\end{eqnarray}
dropping the cumbersome $s$ notation hereafter. The time-dependent spinor modes $f^{u,v}_{\pm {\mathbf k}}(t)$ are solutions to the Dirac equation with a time-dependent mass. Expressing the Dirac equation as a Schr\"odinger equation for the modes
gives
\begin{eqnarray}
    i\dot f_{\mathbf k}(t)=H(t)f_{\mathbf k}(t)
    \ \ .
    \label{eq:det}
\end{eqnarray}
Both $f^u_{\mathbf k}$ and $f^v_{-{\mathbf k}}$ satisfy this Schr\"odinger equation with the Hermitian Hamiltonian
\begin{eqnarray}
   H(t) &=& \gamma^0 ({\mathbf \gamma} \cdot {\mathbf k }+m_f(t)\identity_4)= \pmatrix{-{\mathbf k}\cdot{\mathbf \sigma} & m_f(t)\identity_2\cr m_f(t)\identity_2 & {\mathbf k}\cdot{\mathbf \sigma} }\nonumber \\
      H(t)^2 &=&  \pmatrix{\omega^2_{{\mathbf k}} \identity_2& 0\cr 0 & \omega^2_{{\mathbf k}}\identity_2}
       ,\quad\quad \quad \omega^2_{{\mathbf k}}={\mathbf k}^2+m_f(t)^2
   \label{eq:ham}
\end{eqnarray}
and instantaneous eigenvectors
\begin{eqnarray}
    H(t) u_{{\mathbf k}}&=&\omega_{\mathbf k} (t)u_{{\mathbf k}} \quad \quad \quad \quad  H(t) v_{-{\mathbf k}}=-\omega_{\mathbf k}(t) v_{-{\mathbf k}}
    \nonumber \\
    u^\dagger_{{\mathbf k}}H(t) &=&\omega_{\mathbf k} (t)u^\dagger_{{\mathbf k}} \quad \quad \quad \quad  v^\dagger_{-{\mathbf k}}H(t) =-\omega_{\mathbf k}(t) v^\dagger_{-{\mathbf k}}
    \ \ .
    \label{eq:eigen}
\end{eqnarray}
However, the instantaneous eigenvectors are not necessarily solutions to the time-dependent Dirac equation. 
Using orthogonality and normalizing
\cite{Peskin_1995}
\begin{eqnarray}
    u_{\mathbf k}^{\dagger}u_{\mathbf k}^{}=1
     , \quad\quad v_{-{\mathbf k}}^{\dagger}v_{-{\mathbf k}}^{}=1  , \quad \quad \bar u_{\mathbf k} v_{{\mathbf k}}=0, \quad \quad u_{\mathbf k}^{\dagger}v_{-{\mathbf k}}=0
\end{eqnarray}
although generally, $u_{\mathbf k}^\dagger v_{\mathbf k}\ne 0$.
 This normalization is consistent with (\ref{eq:fk}).

\begin{figure}
        \centering
        \includegraphics[width=0.5\linewidth]{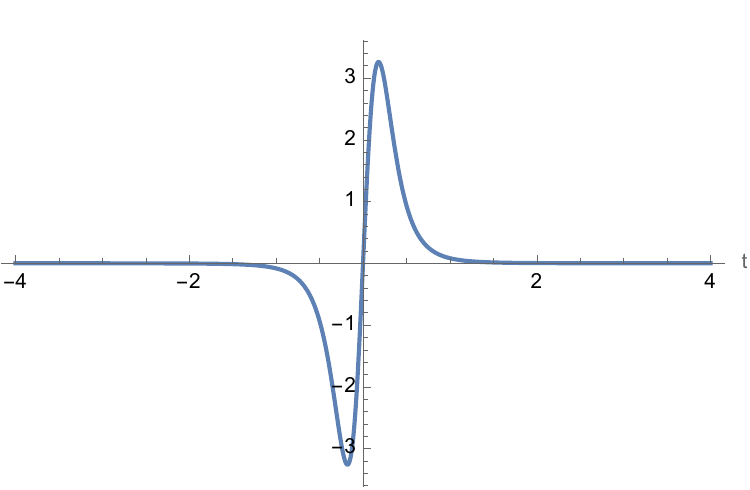}
        \caption{$m_f$ versus $t$ with $g=1/2$, $v_4=1/2$, and $2\pi r_5=0.4$.}
    \label{fig:mf}
\end{figure}

We expand mode solutions to the time-dependent Dirac equation (\ref{eq:det}) in terms of the instantaneous eigenvectors and the Bogoliubov coefficients $\alpha_k$, $\beta_k$ \cite{BirrellDavies1984}:
\begin{eqnarray}
    f^u_{{\mathbf k},s}=\alpha_{\mathbf k} u_{\mathbf k}+\beta_{\mathbf k} v_{-{\mathbf k}}
    \ \ .
\end{eqnarray}
There are two sets of Bogoliubov coefficients, one for $s=1/2$ and one for $s=-1/2$, though
we'll continue to suppress the $s$ index.
Feeding this into (\ref{eq:det}) gives
\begin{eqnarray}
    i\dot \alpha_{\mathbf k} u_{{\mathbf k}}+ i\dot \beta_{\mathbf k} v_{-{\mathbf k}}+ i\alpha_{\mathbf k} \dot u_{{\mathbf k}}+ i \beta_{\mathbf k}\dot v_{-{\mathbf k}}& =& \alpha_{\mathbf k} \omega_{\mathbf k} u_{{\mathbf k}}- \beta_{\mathbf k} \omega_{\mathbf k} v_{-{\mathbf k}}
    \ \ .
    \label{eq:sub}
\end{eqnarray}
Using orthogonality with $u^{\dagger}_{{\mathbf k}}$ isolates $\dot\alpha_{\mathbf k}$ while orthogonality with $v^{\dagger}_{-{\mathbf k}}$ isolates $\dot\beta_{\mathbf k}$:
\begin{eqnarray}
    i\dot \alpha_{\mathbf k} +  i\alpha_{\mathbf k} u^\dagger_{{\mathbf k}}\dot u_{{\mathbf k}}+ i \beta_{\mathbf k}u^\dagger_{{\mathbf k}}\dot v_{-{\mathbf k}}& =& \alpha_{\mathbf k} \omega_{\mathbf k} \nonumber \\
      i\dot \beta_{\mathbf k} +  i\alpha_{\mathbf k} v^\dagger_{-{\mathbf k}}\dot u_{{\mathbf k}}+ i \beta_{\mathbf k} v^\dagger_{-{\mathbf k}}\dot v_{-{\mathbf k}}& =& -\beta_{\mathbf k} \omega_{\mathbf k}
      \ \ .
    \label{eq:abstep}
\end{eqnarray}
To find $v^\dagger_{-{\mathbf k}}\dot u_{{\mathbf k}}$ and similar combinations we multiply the time derivative of $Hu_{\mathbf k}=\omega_{\mathbf k} u_{\mathbf k}$ with $v^\dagger_{-{\mathbf k}}$, which gives
\begin{eqnarray}
  v^\dagger_{-{\mathbf k}}  \dot H(t) u_{{\mathbf k}}+v^\dagger_{-{\mathbf k}}H(t)\dot u_{\mathbf k}&=&\dot \omega_{\mathbf k} v^\dagger_{-{\mathbf k}}u_{{\mathbf k}}+\omega_{\mathbf k} v^\dagger_{-{\mathbf k}} \dot u_{\mathbf k} \nonumber\\
  v^\dagger_{-{\mathbf k}}  \dot H(t) u_{{\mathbf k}}-\omega_{\mathbf k} v^\dagger_{-{\mathbf k}}\dot u_{\mathbf k}&=&\omega_{\mathbf k} v^\dagger_{-{\mathbf k}} \dot u_{\mathbf k} 
  \ \ ,
\end{eqnarray}
having used $v^\dagger_{-{\mathbf k}}H(t)=(H(t)v_{-{\mathbf k}})^\dagger=-\omega_{{\mathbf k}}v^\dagger_{-{\mathbf k}}$ for real $\omega_{{\mathbf k}}$ from eqn.\ (\ref{eq:eigen}).
Solving, we have 
\begin{eqnarray}
     v^\dagger_{-{\mathbf k}} \dot u_{\mathbf k} =\frac{1}{2\omega_{\mathbf k}}v^\dagger_{-{\mathbf k}}  \dot H(t) u_{{\mathbf k}}
     \ \ .
\end{eqnarray}
We can push this a bit further with an operator related to the helicity operator
\begin{equation}
    h_{\mathbf k}={\mathbf k}\cdot S=\frac{1}{2}\pmatrix{{\mathbf k}\cdot{\mathbf \sigma} & 0\cr 0 &{\mathbf k}\cdot{\mathbf \sigma} }
    \ \ .
\end{equation}
Reviving the $s$ index, we note that
\begin{eqnarray}
    h_{\mathbf k}u_{{\mathbf k},s}=sk u_{{\mathbf k},s}\ , \quad\quad\quad
    h_{\mathbf k}v_{{\mathbf k},s}=-sk v_{{\mathbf k},s}
    \ \ ,
    \label{eq:hs}
\end{eqnarray}
where $s=\pm \frac{1}{2}$ and by $k$ we mean the magnitude of the $3$-momentum, $k=|{\mathbf k}|$.
We'll use this and reexpress
\begin{eqnarray}
    H(t)=2\bar \gamma h_{\mathbf k} +\gamma^0 m_f
    \ \ .
\end{eqnarray}
We simplify the following:
\begin{eqnarray}
v^\dagger_{-{\mathbf k}}  \dot H(t) u_{{\mathbf k}}&=&\dot m_fv^\dagger_{-{\mathbf k}} \gamma^0u_{{\mathbf k}}
     =\frac{\dot m_f}{\omega_{\mathbf k}}v^\dagger_{-{\mathbf k}} \gamma^0 H u_{{\mathbf k}}
     =\frac{\dot m_f}{\omega_{\mathbf k}}v^\dagger_{-{\mathbf k},s} 
     \gamma^02\bar \gamma h_{\mathbf k} u_{{\mathbf k},s}\nonumber \\
     &=&
     \frac{2sk\dot m_f}{\omega_{\mathbf k}}v^\dagger_{-{\mathbf k},s} 
     \gamma^0\bar \gamma u_{{\mathbf k},s}=
     \frac{2sk\dot m_f}{\omega_{\mathbf k}}v^\dagger_{-{\mathbf k},s} 
      v_{-{\mathbf k},s}=
\pm \frac{k\dot m_f}{\omega_k}
\ \ .
\end{eqnarray}
We used $2s=\pm 1$ and
\begin{eqnarray}
    v_{-{\mathbf k},s}=\gamma^0\bar\gamma u_{{\mathbf k},s}
    \ \ ,
\end{eqnarray}
which is consistent with eqn.\ (\ref{eq:eigen}).
Again, the upper sign corresponds to $s=1/2$ while the lower sign corresponds to $s=-1/2$. 

The combinations we need are
\begin{eqnarray}
     v^\dagger_{-{\mathbf k}} \dot u_{\mathbf k} &=&\frac{1}{2\omega_{\mathbf k}}v^\dagger_{-{\mathbf k}}  \dot H(t) u_{{\mathbf k}}
     =\pm\frac{k\dot m_f}{2\omega^2_{\mathbf k}}\nonumber \\
     u^\dagger_{{\mathbf k}} \dot v_{-{\mathbf k}} &=&-\frac{1}{2\omega_{\mathbf k}}u^\dagger_{{\mathbf k}}  \dot H(t) v_{-{\mathbf k}}=\mp\frac{ k\dot m_f}{2 \omega^2_{\mathbf k}} \nonumber \\
     v^\dagger_{-{\mathbf k}} \dot v_{-{\mathbf k}} &=&u^\dagger_{{\mathbf k}} \dot u_{\mathbf k} =0
     \ \ .
\end{eqnarray}
The final line amounts to a gauge choice for an arbitrary phase.
Using this in (\ref{eq:abstep})
\begin{eqnarray}
    \dot \alpha_{\mathbf k} & =&-i  \omega_{\mathbf k} \alpha_{\mathbf k} \pm \beta_{\mathbf k}k\frac{\dot m_f}{2\omega^2_{\mathbf k}}\nonumber \\
       \dot \beta_{\mathbf k} & =& i \omega_{\mathbf k} \beta_{\mathbf k}   \mp \alpha_{\mathbf k} k\frac{\dot m_f}{2\omega^2_{\mathbf k}}
       \ \ .
    \label{eq:abstep2}
\end{eqnarray}
These are often easier to integrate numerically but we could also
shift the Bogoliubov coefficients to separate out the fast oscillations 
\begin{eqnarray}
    \alpha_{\mathbf k}&=&\tilde \alpha_{\mathbf k} e^{-i\int^t \omega(t^\prime)dt^\prime}
    \nonumber \\
    \beta_{\mathbf k}&= &\tilde \beta_{\mathbf k} e^{i\int^t \omega(t^\prime)dt^\prime}
    \ \ .
\end{eqnarray}
Using $\dot m_f/\omega_{\mathbf k}=\dot \omega_{\mathbf k}/m_f$,
we have the general result for the evolution of the Bogoliubov coefficients for a fermion with a time-dependent mass:
\begin{mdframed}
\begin{eqnarray}
    \dot {\tilde \alpha}_{\mathbf k} & =& \pm {\tilde \beta}_{\mathbf k} \left (\frac{k}{m_f}\right )\frac{\dot \omega_{\mathbf k} }{2\omega_{\mathbf k}}e^{2i\int^t\omega(t^\prime)dt^\prime }\nonumber \\
    \dot {\tilde \beta}_{\mathbf k} & =&  \mp {\tilde \alpha}_{\mathbf k}\left ( \frac{k}{m_f}\right )\frac{\dot \omega_{\mathbf k}}{2\omega_{\mathbf k}} e^{-2i\int^t\omega(t^\prime)dt^\prime} \ \ ,
    \label{eq:bogf}
\end{eqnarray}
\end{mdframed}
with $|\alpha_{\mathbf k}|^2+|\beta_{\mathbf k}|^2=1$.
Again, the upper sign is for $s=1/2$ while the lower sign is for $s=-1/2$.

As compared to the more familiar scalar equations for which $|\alpha_{\mathbf k}|^2-|\beta_{\mathbf k}|^2=1$,
\begin{eqnarray}
    \dot {\tilde \alpha}_{\mathbf k} & =& {\tilde \beta}_{\mathbf k} \frac{\dot \omega_{\mathbf k}}{2\omega_{\mathbf k}}e^{2i\int^t\omega(t^\prime)dt^\prime }\nonumber \\
    \dot {\tilde \beta}_{\mathbf k} & =&  {\tilde \alpha}_{\mathbf k} \frac{\dot \omega_{\mathbf k}}{2\omega_{\mathbf k}} e^{-2i\int^t\omega(t^\prime)dt^\prime} 
    \ \ ,
    \label{eq:bogscalar}
\end{eqnarray}
the fermion equations have a  relative sign difference and a factor of $k/m_f$. 

\begin{figure}
    \centering
    \includegraphics[width=0.45\linewidth]{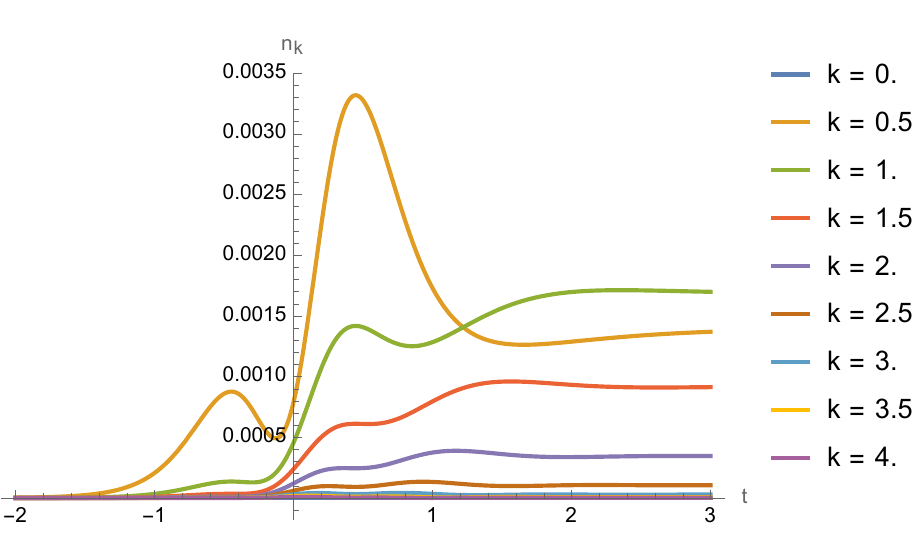}
    \includegraphics[width=0.45\linewidth]{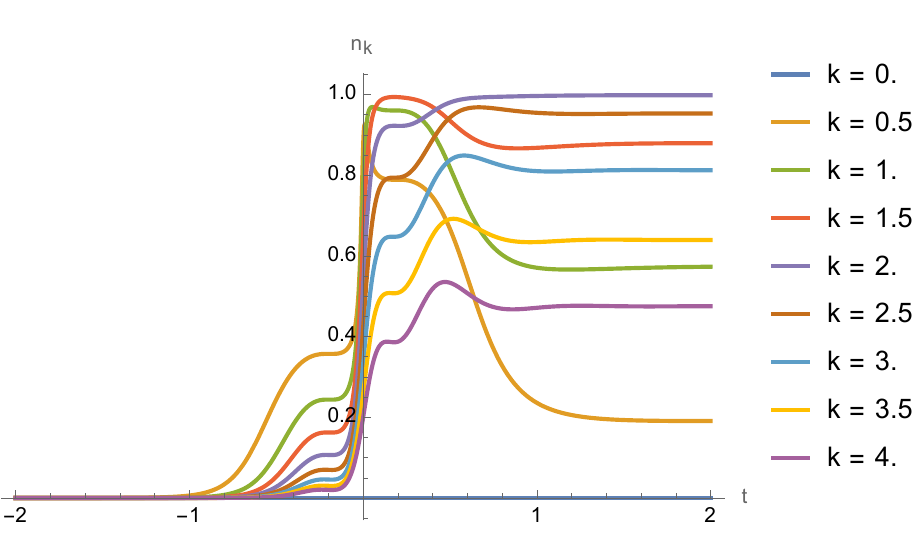}
    \caption{The particle number density for a given spin as a function of time and for a range of modes. The parameters are $g=1/2$, $v_4=1/2$ with $2\pi r_5=1$ on the left and $2\pi r_5=0.4$ on the right.}
    \label{fig:nk}
\end{figure}

In the far past, the instantaneous eigenmodes are solutions to the massless Dirac equation and correspond to a well-defined vacuum. In the far future, the instantaneous eigenmodes are again solutions to the massless Dirac equation and the Bogoliubov coefficients 
account for the mismatch in early- and late-time modes. The number density 
of particles created is given by $n_{\mathbf k}=|\beta_{\mathbf k}|^2$ at late times.
We show in figure\ \ref{fig:nk} the numerically integrated
$n_{\mathbf k}=|\beta_{\mathbf k}|^2$, which we expect to always be less than 1 due to fermi statistics. Notice that there is a large spike around $t=0$ when $\dot m_f/\omega_{\mathbf k}^2$ is a maximum. This suggests a nonadiabatic burst of particle production. 
We should get this burst twice per orbit as the brane hits two walls for $N$ orbits. The accumulated phase could delay the peak to just after $t=0$ if the peak is not sharp. 

The energy density is
\begin{equation}
    \rho=g_s\int d^3k\ \omega_{\mathbf k}n_{\mathbf k}
    \ \ ,
\end{equation}
where $g_2=2$ accounts for the two spin states. As we have assumed that the Klein bottle geometry has been stabilized, the only source for this energy production is a reduction in the brane's kinetic energy, which will decrease with each new instance of particle production. Eventually the brane will come to rest and particle production will end. The final mass of the brane fermion will depend on the resting place of the braneworld in the Klein bottle.

\subsection{Implications for Cosmology}
\label{sec:baryo}

We have seen that there are bursts of particle production as the brane passes through the condensate wall. 
To place this in the larger cosmological context, we consider the conditions needed to create the universe's observed matter asymmetry.
The ${cp}$ violating phases in the standard model Lagrangian are too small to meet the demands of baryogenesis, which requires an asymmetry today of 
\begin{eqnarray}
    \eta=\frac{n_B-n_{\bar B}}{s}\sim 10^{-1}\epsilon\sim 8.6 \times 10^{-11}
    \ \ ,
\end{eqnarray}
where the numerator is the difference between the baryon number density and the antibaryon number density and the denominator is the entropy density of the universe \cite{KolbTurner1990}. The ${cp}$ violation is quantified in the parameter $\epsilon$, which can be related to elements in the fermion mass matrices, for instance.

The Klein bottle condensate can contribute new sources of ${cp}$ violation and thereby revive baryogenesis. We have set up all of the ingredients for leptogenesis, which customarily generates a lepton asymmetry as a byproduct of out-of-equilibrium decays of a heavy right-handed neutrino. 
The condensate potentially rewrites leptogenesis by providing a spontaneously ${cp}$ violating Majorana mass matrix for multiple right-handed neutrinos,
\begin{eqnarray}
    -\frac{1}{2}\bar {\mathbf f}^{c} M {\mathbf f} (\bar \Psi i\bar \Gamma \Psi) +h.c.
\end{eqnarray} 
where $M$ is a matrix built out of complex couplings to the condensate
$M_{ij}=y_{ij}\langle {\mathbf \Psi}\bar i \bar\Gamma  {\mathbf \Psi}\rangle$ for multiple bulk fermions.
The Majorana mass matrix above is a module in leptogenesis that could also include Dirac mass terms as well as other non-${cp}$ violating mass terms. It is worth noting that the condensate can also introduce ${cp}$ violation in the decays of the heavy neutrino that lead to leptogenesis.
The standard model can then do the rest, transforming lepton number violation into baryon number violation with $B-L$ conserving sphaleron transitions active at $T> 100 $ GeV.  

Typical mass ranges for leptogenesis can be as high as 
\begin{eqnarray}
    M_L\sim 10^9-10^{14} \ {\rm GeV}
    \ \ .\label{eq:Mgen}
\end{eqnarray}
As the brane sweeps through the wall, there would be bursts of production of heavy right-handed neutrinos, whose mass peaks at around
\begin{eqnarray}
    M_L\sim g\frac{1}{r_5^5}
\end{eqnarray}
with the units of $[g]\sim [r_5]^{4}$, which translates to a size for the space in the range
\begin{eqnarray}
    r_5\sim 10^{-23}-10^{-28} \ {\rm cm}
    \ \ .
    \label{eq:R5}
\end{eqnarray}
If the brane comes to rest near the bottom of the wall, then these same neutrinos contribute to dark matter, which in a CDM WIMP scenario typically have masses in the range
\begin{eqnarray}
    M\sim 1\ {\rm GeV}-10\ {\rm TeV}
\end{eqnarray}
if the brane resting place is very near the origin of the wall, $x_{4b}\sim 10^{-6}-10^{-14} \ r_5$. Equivalently, the brane could come to rest between walls, roughly $x_{4b}\sim 10^{2}-10^{3} \ r_5$.
Our ambitions do not extend to naturalness, technical or otherwise.

We mention that through the dependence on the brane location, the magnitude of the ${cp}$ violation varies as the brane moves and the velocity changes. The ${cp}$ violation is also dynamical on cosmological timescales if the size of the internal dimensions is dynamical, with the maximum of the wall function scaling as $r_5^{-5}$. That flexibility allows for a landscape of possibilities that would impact the early universe yet evade detection in the low-energy physics of today. It would be intriguing to apply this novel topological mechanism for $cp$ violation to other important quandaries, such as the strong ${cp}$ problem.

While we leave the complex details of this novel baryogenesis proposal for further study, we conclude by emphasizing the key ingredients of the scenario: 

$\indent \bullet$ Non-equilibrium particle production due to brane motion through the wall

$\indent \bullet$ Spontaneous ${cp}$ violation in the neutrino Majorana mass matrix

$\indent \bullet$ Early universe, heavy, fermion for leptogenesis

$\indent \bullet$ Lighter, but still heavy, neutrinos as CDM candidates.

The braneworld proves valuable since the scenario meets all three of Sakharov's conditions for baryogenesis. However, if baryogenesis can transpire in the full higher-dimensional spacetime, we could also exploit the explicit breaking of $(5+1)$-dimensional ${\C\PJ}$ in the Klein bottle.

\bigskip
\goodbreak
\centerline{\bf Acknowledgements}
\noindent
BG is supported in part by DOE award DE-SC0011941. DK is supported by U.S.\ National Science Foundation grant PHY-2412480. JL is supported in part by the Tow Foundation.
MP is supported in part by NSF grant PHY-2210349.

\printbibliography

\end{document}